\documentclass[final]{aipproc}
\usepackage{graphicx}
\layoutstyle{6x9}

\begin{document}

\title{Constraining dark energy interacting models with WMAP}

\author{Germ\'{a}n Olivares}
{address={Departamento de F\'{i}sica, Universidad Aut\'{o}noma de
Barcelona, Spain}}
\author{Fernando Atrio-Barandela}
{address={Departamento de F\'{i}sica Te\'{o}rica, Universidad de Salamanca,
 Spain}}
\author{Diego Pav\'{o}n}
{address={Departamento de F\'{i}sica, Universidad Aut\'{o}noma de
Barcelona, Spain}}

\begin{abstract}
We determine the range of parameter space of an interacting
quintessence (IQ) model that best fits the luminosity distance of
type Ia supernovae data and the recent WMAP measurements of Cosmic
Microwave Background temperature anisotropies. Models in which
quintessence decays into dark matter provide a clean explanation
for the coincidence problem. We focus on cosmological models of
zero spatial curvature. We show that if the dark energy (DE)
decays into cold dark matter (CDM) at a rate that brings the ratio
of matter to dark energy constant at late times, the supernovae
data are not sufficient to constrain the interaction parameter.
On the contrary, WMAP data constrain it to be smaller than
$c^2 < 10^{-2}$ at the $3\sigma$ level. Accurate measurements of the
Hubble constant and the dark energy density, independent of the
CMB data, would support/disprove this set of models.
\end{abstract}

\maketitle

\section{Introduction}
Recent observational data suggest the Universe has entered a
period of accelerated expansion \cite{SNIa}. The $\Lambda$CDM
model is the simplest model that best fits the supernovae
luminosity distance and CMB anisotropy spectrum data
\cite{SNIa,WMAP}.  Though it looks a very serious candidate it
raises two main theoretical problems. First, it assumes the energy
density of the vacuum but according to quantum field theory it
should be larger than observed by $120$ orders of magnitude.
Secondly, there is the coincidence problem, namely: {\em why the
energy densities of matter and dark energy -scaling  so
differently with expansion- happen to be of the same order
precisely today?} To solve the latter problem models featuring an
interaction between quintessence and cold dark matter has been
proposed \cite{iqm0,iqm}.

\section{The interacting quintessence model}
\begin{figure}
    \includegraphics[height=.3\textheight]{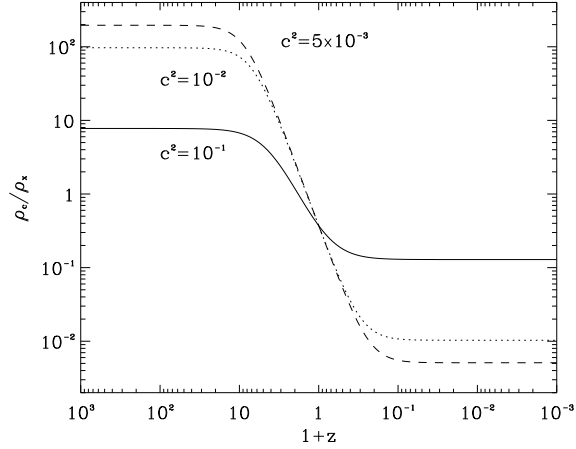}
    \caption{Evolution of the energy density ratio $r=\rho_{cdm}/\rho_{x}$
from an unstable maximum towards a stable minimum (at late times)
for different values of $c^2$. As current value we take $r_{0} =
0.42$.}
\label{fig:ratio}
\end{figure}

To account for the coincidence problem Zimdahl et al. \cite{iqm0}
postulated an interaction between cold dark matter and dark energy
(a quintessence scalar field).
\\
\begin{equation}\label{cont_eqs}
\dot\rho_{x} + 3H(1+w_x)\rho_{x} = -\delta \,  , \qquad \quad
\dot\rho_{cdm} + 3H\rho_{cdm} = \delta .
\end{equation}
\\
This brings the
ratio between their energy densities, $r \equiv%
\rho_{cdm}/\rho_{x}$, to a stable constant value at late times.
The interaction term, $\delta$, may depend on $H$, $\rho_{x}$ and
$\rho_{cdm}$. A simple ansatz is
$\delta=3Hc^2(\rho_x+\rho_{cdm})$,
 where the constant $c^2$ is the interaction parameter. From Eqs.
 (\ref{cont_eqs}) the aforesaid ratio
 is constrained to evolve between two stationary values,  $r_{\pm} =%
-\frac{1+w_x}{c^2} \pm \sqrt{\frac{w_x^2}{4c^4}+\frac{w_x}{c^2}}$,
where $r_{+}$ denotes the unstable value and  $r_{-}$ the stable
one -see Ref. \cite{iqm} for details. As seen from Fig.
\ref{fig:ratio} the stronger the interaction, the smaller
$\rho_{cdm}$ in the past. Obviously, such a dissimilar evolution
of $\rho_{cdm}$ as compared to the $\Lambda$CDM model, gives rise
to a different expansion rate and a different evolution of the
metric perturbations. The former modifies the luminosity distance
of SNIa. The latter should be observed in the CMB anisotropy
spectrum. Below we  constrain the IQ model with the SNIa data of
Riess et al. \cite{SNIa} and the CMB anisotropy data collected by
the WMAP satellite \cite{WMAP}.
\\

\section{Observational constraints}
\begin{figure}
\includegraphics[height=.3\textheight]{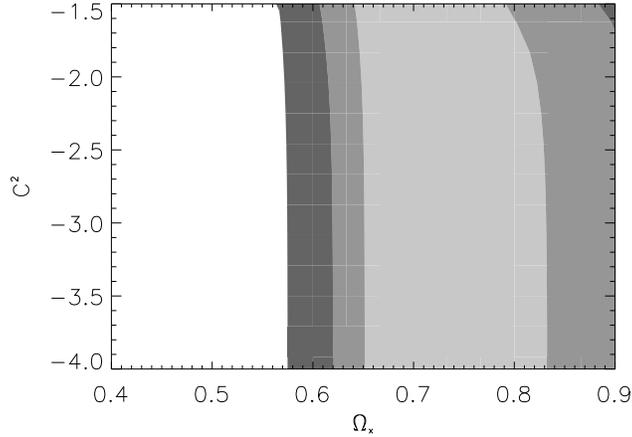}
\caption{Joint confidence intervals at $68\%$, $95\%$ and $99.9\%$
C.L. of the IQ model fitted to the ``gold" sample of SNIa data of
Riess et al. \cite{SNIa}. For convenience, the $c^{2}$ axis is
represented using a logarithmic scale and it has been cut to
$c^{2}\leq 10^{-4}$, though models with $c^{2} = 0$ have been
included in the analysis.} \label{fig:SNIa}
\end{figure}

To constrain the IQ model with the SNIa data we assume the
following priors: a spatially flat model ($\Omega_k=0$) and  $-1 <
w_x < -0.6$. As seen from Fig. \ref{fig:SNIa} the lack of data for
redshifts $z > 1.8$ does not constrain $c^2$ \cite{olivares}.
\\

\begin{figure}
    \includegraphics[height=.3\textheight]{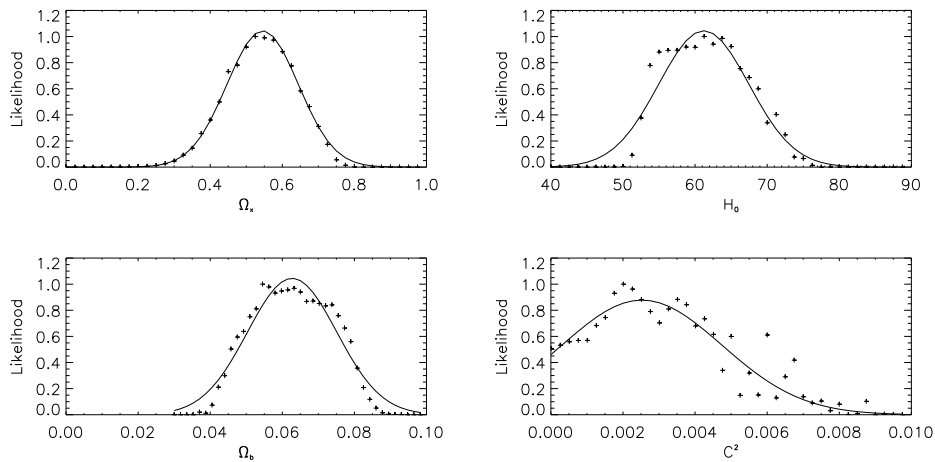}
    \caption{Crosses represent the likelihood computed from
    $50,000$ models, marginalized over all parameters but one.
    The solid line represents the best Gaussian fit.}
    \label{fig:like2d}
\end{figure}

To constrain the model with the WMAP data we adopt a
phenomenological point of view and impose that the interaction
exists along the streamlines of the fluids (dark energy and CDM),
i.e.,
\\
\begin{equation}\label{cont_first}
\begin{array}{ccc}
u^{\nu}\nabla_{\mu}T^{\mu}_{\nu}{}_{(cdm)}& = & -u^{\mu}\nabla_{\mu}\rho_{cdm}-
3 H (\rho_{cdm}+P_{cdm})=-\delta \, .\\
\end{array}
\end{equation}
\\
The net momentum transfer between CDM and dark energy alongside
the accelerated expansion, limits the growth of the CDM
perturbations. We implement the perturbations equations
(\ref{cont_first}) together with
$\nabla_{\mu}T^{\mu}_{\nu}{}_{(x)}=-\nabla_{\mu}T^{\mu}_{\nu}{}_{(cdm)}$
in the CMBfast code \cite{cmbfast} and assume a flat model with
($\Omega_k=0$), universes older than $12$ Gyrs, BBN, and $-1 < w_x
< 0.6$. By means of a Monte Carlo Markov chain  we reconstruct the
surface of the likelihood in the  7D-space parameter  defined by
the cosmological parameters $c^2, \Omega_x, w_x, \Omega_b, H_0,
 n_s, A$.
\begin{table}
\begin{tabular}{lcc}
\hline
   \tablehead{1}{c}{b}{Parameter}
  & \tablehead{1}{r}{b}{IQ model}
  & \tablehead{1}{r}{b}{$\Lambda$CDM model \tablenote{WMAP team results \cite{WMAP}.}}\\
\hline
Interaction parameter, $c^2$& $< 5\times 10^{-3}$\tablenote{Upper limit at $1\sigma$ C.L.}  &
 $0$\tablenote{This is a prior of the $\Lambda$CDM model.}\\
CDM density/critical density, $\Omega_c$ & $0.4\pm 0.1 $ & $0.29\pm 0.07 $\\
Baryon density/critical density, $\Omega_b$ & $0.063\pm 0.012 $ & $0.047\pm 0.006 $\\
Hubble constant, $h_0$ & $0.61\pm 0.06 $ & $0.72\pm 0.05 $\\
Spectral tilt, $n_s$ & $0.96\pm0.03$ & $0.99\pm 0.04$\\
Amplitude, $A$ & $0.9\pm 0.1$ & $0.9\pm 0.1$ \\
\hline
\end{tabular}
\caption{Mean values and $1\sigma$ C.L. of the parameters of the
IQ and $\Lambda$ CDM models.}
\label{tab:a}
\end{table}
Table \ref{tab:a} summarizes our results. Due to the interaction
the CDM redshift is smaller than in the $\Lambda$CDM model,
therefore a higher value of $\Omega_{cdm}$ is needed to have
enough gravitational potential at matter-radiation decoupling.
Moreover, in order not to alter the height of the acoustic peaks
the $\Omega_{b}$ value increases. Otherwise, the small mean value
of $H_{0}$ would suppress them. The likelihood shows clearly that
the interaction parameter $c^2$ is compatible with the WMAP data
(see Fig. \ref{fig:like2d}). Nevertheless, non-interacting models
are also allowed, unless the amplitude, $A$, is not a parameter
(see Ref. \cite{olivares}). Thus, $c^2$ is bounded from above. At
$1\sigma$ confidence level  $c^2<5\times 10^{-3}$. The spectral
tilt, $n_s$, and the amplitude of the perturbations at the horizon
size, $A$, are in perfect agreement with those of the WMAP team
\cite{WMAP}.

\section{Conclusions}
We have presented an interacting cosmological model that solves
the coincidence problem. Despite the seemingly odd results of Tab.
\ref{tab:a} it is in perfect agreement with SNIa data \cite{SNIa}
and CMB data \cite{WMAP}. Moreover, the lack of power at the lower
multipoles becomes a less serious problem. Indeed, because of the
smaller redshift evolution of CDM energy density the gravitational
potential evolves slower than in the $\Lambda$CDM model. Thus, a
smaller Integrated Sachs-Wolfe effect produces less power in the
CMB anisotropy spectrum. A next generation of accurate
measurements of the Hubble parameter and the dark energy density,
independent of the CMB data, are needed to support/disprove this
kind of models.

\begin{theacknowledgments}
We thank the organizers of the XXVIIIth  edition of the
``Encuentros Relativistas Espa\~{n}oles" for this opportunity.
This research was partly supported by the Spanish Ministry of
Science and Technology under Grants BFM2003-06033, BFM2000-1322
 and the Junta de Castilla y Le\'{o}n (projects SA002/03, SA010C05).
\end{theacknowledgments}

\end{document}